\documentclass[aps, prl, showpacs, twocolumn, amsfonts, amsmath, amssymb, superscriptaddress, floatfix]{revtex4-1}
\usepackage[colorlinks,bookmarks=false,citecolor=blue,linkcolor=red,urlcolor=black]{hyperref}

\usepackage[T1]{fontenc}
\usepackage[latin9]{inputenc}
\usepackage{color}
\usepackage{graphicx}
\usepackage{ulem}
\usepackage{epsfig}
\def\be{\begin{equation}}
\def\ee{\end{equation}}
\def\bea{\begin{eqnarray}}
\def\eea{\end{eqnarray}}

\begin{document}

\title{Time crystal behavior of excited eigenstates}

\author{Andrzej Syrwid} 
\affiliation{
Instytut Fizyki imienia Mariana Smoluchowskiego, 
Uniwersytet Jagiello\'nski, ulica Profesora Stanis\l{}awa \L{}ojasiewicza 11, PL-30-348 Krak\'ow, Poland}

\author{Jakub Zakrzewski} 
\affiliation{
Instytut Fizyki imienia Mariana Smoluchowskiego, 
Uniwersytet Jagiello\'nski, ulica Profesora Stanis\l{}awa \L{}ojasiewicza 11, PL-30-348 Krak\'ow, Poland}
\affiliation{Mark Kac Complex Systems Research Center, Uniwersytet Jagiello\'nski, ulica Profesora Stanis\l{}awa \L{}ojasiewicza 11, PL-30-348 Krak\'ow, Poland
}

\author{Krzysztof Sacha} 
\affiliation{
Instytut Fizyki imienia Mariana Smoluchowskiego, 
Uniwersytet Jagiello\'nski, ulica Profesora Stanis\l{}awa \L{}ojasiewicza 11, PL-30-348 Krak\'ow, Poland}
\affiliation{Mark Kac Complex Systems Research Center, Uniwersytet Jagiello\'nski, ulica Profesora Stanis\l{}awa \L{}ojasiewicza 11, PL-30-348 Krak\'ow, Poland
}


\begin{abstract}
In analogy to spontaneous breaking of continuous space translation symmetry in the process of space crystal formation, it was proposed that spontaneous breaking of continuous time translation symmetry could lead to time crystal formation. In other words, a time-independent system prepared in the energy ground state is expected to reveal periodic motion under infinitely weak perturbation. In the case of the system proposed originally by Frank Wilczek, spontaneous breaking of time translation symmetry can not be observed if one starts with the ground state. We point out that the symmetry breaking can take place if the system is prepared in an excited eigenstate. The latter can be realized experimentally in ultra-cold atomic gases. We simulate the process of the spontaneous symmetry breaking due to measurements of particle positions and analyze the lifetime of the resulting symmetry broken state.
\end{abstract}

\maketitle

Hamiltonians of condensed matter systems are invariant under translation of all particles by the same vector in space and so are the eigenstates. Consequently probability density for detection of a single particle must be uniform in space if a system is prepared in the ground state or any other eigenstate. Space crystals emerge due to spontaneous symmetry breaking that can be induced by an external perturbation or by, e.g., measurements of particle positions. If the single particle probability density is uniform but the density-density correlation function reveals periodic behavior, measurement of a position of a particle breaks the continuous space translation symmetry and the probability density for the detection of a next particle shows crystalline structure \cite{kaplan89,koma93}. In the thermodynamic limit, the lifetime of the symmetry broken state goes to infinity and the stable space crystal is formed.

Similar phenomenon was postulated to exist in the time domain \cite{wilczek12}. A spontaneous breaking of the continuous time translation symmetry  in the ground state of the model system was suggested to lead to a periodic motion of a nonuniform density. Soon the other experiment involving trapped ions in a ring  \cite{li12} was proposed. However, at the same time the original proposition has been put in question
 \cite{bruno13,bruno13a}. While the discussion interestingly evolved \cite{li12a,wilczek13,wilczek13a,chernodub12,volovik13,mendonca13,robicheaux15}  strong arguments have been presented \cite{bruno13b,nozieres13,watanabe14} against the existence of time crystals. 
%Yet some argue that time crystals actually exist  already \cite{wu16}. 
The proposals, nevertheless, became inspiring and triggered a new field of research. It turns out that, by analogy to condensed matter physics, where space periodic potentials allow for modeling of space crystals, periodically driven systems can model crystalline behavior in time \cite{sacha14}. Anderson localization and Mott insulator phase in the time domain can be realized \cite{sacha15,sacha16,Delande_Morales_Sacha} and spontaneous 
breaking of discrete time translation symmetry to another discrete symmetry can be investigated \cite{sacha14}. The latter phenomenon, termed {\it discrete time crystal}, was recently observed in two experiments \cite{zhang16,choi16} following independent theoretical suggestions \cite{khemani16,else16,else16a,keyserlingk16a,keyserlingk16b,keyserlingk16,yao16}. Those propositions, contrary to the original work \cite{sacha14}, relied on the many-body localization phenomenon stabilizing the proposed phase in the presence of the disorder.

The original idea of forming the time crystal by spontaneous symmetry breaking \cite{wilczek12} negated by others \cite{bruno13b,nozieres13,watanabe14}  considered the ground state of the system studied.  However, all eigenstates of a time-independent Hamiltonian system trivially possess a continuous time translation symmetry being stationary states. In the present work we point out that for some of those eigenstates, even highly excited ones, a spontaneous breaking of  this continuous time translation symmetry to a discrete one can be realized even  in the limit of an infinite number of particles. Such excited eigenstates can be prepared in ultra-cold atoms laboratories and effects due to the spontaneous symmetry breaking can be observed experimentally. Let us stress that 
this proposition is orthogonal to recent works on discrete or Floquet time crystals \cite{sacha14,zhang16,choi16,khemani16,else16,else16a,keyserlingk16a,keyserlingk16b,keyserlingk16,yao16} and relies on the continuous time translational symmetry breaking.

Before we proceed, it is necessary to clarify the definition of time crystals. In the space crystal case, the crystalline structure is defined as a periodic behavior of the probability density for the measurement of particle positions in space at a fixed moment of time. In the case of time crystals the role of space and time are exchanged. That is, we should fix the position coordinate in the configuration space and look for periodic behavior of the detection probability versus time \cite{li12,li12a,robicheaux15}. In other words, a detector is placed at a certain position and periodic clicking of the detector in time is expected. This definition of time crystals does not require the thermodynamic limit that is usually considered in condensed matter physics as in Ref.~\cite{watanabe14}, i.e., we do not need the volume of the system $V\rightarrow\infty$ because we do not need to ensure periodic behavior (or any other behavior) in space. However, in order to deal with a time crystal we have to ensure that once the symmetry is broken, the 
quantum evolution reveals 
periodic behavior forever. The latter can be fulfilled in  Wilczek's model if the number of particles $N\rightarrow\infty$ but $N$ times the coupling constant $g_0$ is fixed which constitutes the standard mean field limit in ultra-cold atomic gases. In order to observe time crystal behavior of an excited eigenstate, thermal cloud has to be sufficiently eliminated to prevent dissipation. That again is possible in ultra-cold atoms.

We consider Wilczek's model \cite{wilczek12} of time crystals where $N$ bosons move on a one-dimensional Aharonov-Bohm ring of unit length and interact via an attractive contact potential. The quantum  Hamiltonian of the system reads, 
\be
H=\sum_{i=1}^N\frac{(p_i-\alpha)^2}{2}+\frac{g_0}{2}\sum_{i\ne j}\delta(x_i-x_j),
\label{h}
\ee
where $\alpha$ is the constant parameter that can be interpreted as a magnetic flux through the ring and $g_0<0$ determines the strength of the attractive interactions between particles. We assume $m=\hbar=1$. 

Let us begin with the analysis of the $\alpha=0$ case. In order to find the ground state for bosons one can apply the mean field approach where the $N$-body ground state can be approximated by a product state $\psi=\prod_{i=1}^N\phi_0(x_i)$ with $\phi_0$ being the lowest energy solution of the Gross-Pitaevskii equation,
\be
\left(-\frac12  \partial_x^2+g_0(N-1)|\phi_0|^2\right)\phi_0=\mu\phi_0,
\ee
$\mu$ being the chemical potential. For small negative values of $g_0(N-1)$, the ground state solution of the Gross-Pitaevskii equation on the ring is uniform, i.e. $\phi_0=1$. However, when $g_0(N-1)<-\pi^2$, it becomes energetically favorable to keep particles close to each other and a non-uniform solution becomes the ground state \cite{carr00}. In the case of periodic boundary conditions we consider here, $\phi_0$ is given by the Jacobi elliptic function which tend to the well known bright soliton shape $\phi_0\propto 1/\cosh[g_0(N-1)(x-x_{\rm CM})/2]$ for $g_0N\rightarrow-\infty$. The center of mass (CM) coordinate $x_{\rm CM}$ is a parameter. Its value can be chosen arbitrarily  which indicates that the mean field approach predicts breaking of space translation symmetry contrary to the exact many-body description where the eigenstates must be translationally invariant. Note that the Gross-Pitaevskii equation does not depend on $g_0$ and $N$ separately but on the product $g_0(N-1)$.  In the 
following we consider the limit where $N\rightarrow\infty$, $g_0\rightarrow 0$ but $g_0(N-1)=$constant. In that limit, the mean field predictions remain unchanged.

Now, let us return to the $\alpha\ne 0$ case. If we switch to the CM coordinate frame, the Hamiltonian (\ref{h}) becomes
\be
H=\frac{(P-N\alpha)^2}{2N}+\tilde H(\tilde x_i,\tilde p_i),
\label{hcm}
\ee
where $P$ is the center of mass momentum and $\tilde H(\tilde x_i,\tilde p_i)$ is the Hamiltonian expressed in relative degrees of freedom. Eigenstates of the system are determined by an independent choice of the CM momentum (which is quantized, $P_j=2\pi j$ with integer $j$) and the relative degrees of freedom quantum numbers. The ground state corresponds to the minimal value of the first term in the right-hand-side of (\ref{hcm}), i.e., to
\be
\frac{\partial H}{\partial P_j}=2\pi\frac{j}{N}-\alpha\approx 0,
\ee
which can be chosen to be exactly zero in the limit of $N\rightarrow\infty$. It implies that there is no CM motion in the ground state for a large particle number. Thus, even if a spontaneous localization of the CM took place, we would not observe the spontaneous breaking of the time translation symmetry because the localized CM would not move. This constitutes a simple argument why  Wilczek's idea in its original version does not work.  However, there is the CM probability current for, e.g., $P_N=2\pi N$ even for large $N$ if $\alpha$ is chosen so that
\be
\frac{\partial H}{\partial P_N}=2\pi-\alpha\ne 0.
\label{current}
\ee
The eigenstate corresponding to $P_N$ is not the ground state because one can always choose such $P_j$ that leads to a lower energy of the CM degree of freedom. If, however, the excited eigenstate with $P_N$ can be experimentally prepared, then analysis of possibility of spontaneous time translation symmetry breaking for this state is not the theoretical issue only. One can imagine an experiment where the system is initially prepared in the ground state for $\alpha=2\pi$ which corresponds to $P_N$. Next, $\alpha$ is switched to zero that makes the initial state with $P_N$ an excited eigenstate of the new Hamiltonian with the CM probability current equal to $2\pi$, cf. (\ref{current}). In the following we assume $\alpha=0$ and that the initial state $|\psi_0\rangle$ corresponds to the lowest energy eigenstate in the subspace with the total momentum $P_N$ and investigate if measurement of particles positions breaks the continuous time translation symmetry and pushes the system to 
periodic motion that can live 
infinitely long in the limit of $N\rightarrow\infty$ \cite{li12,li12a,sacha14,robicheaux15}.
\begin{figure*}
 \includegraphics[width=1.0\linewidth]{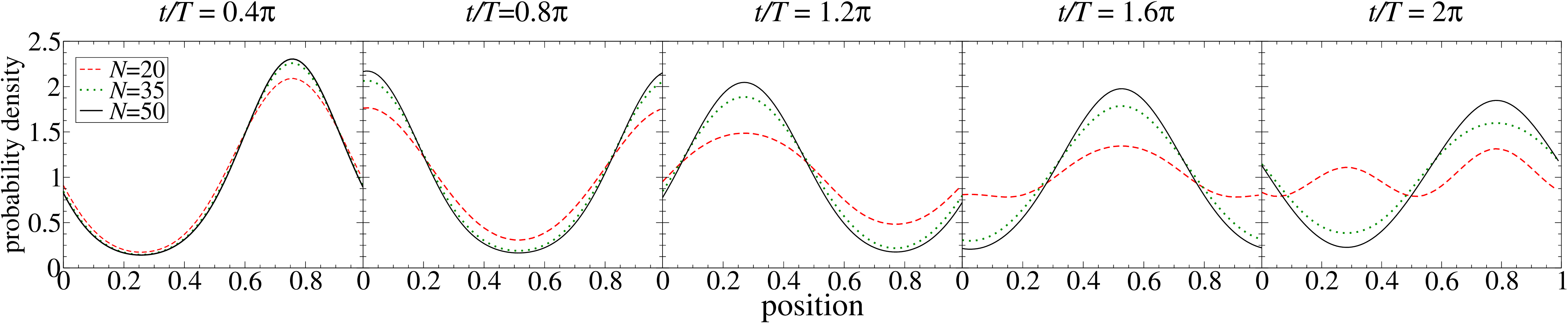}
\caption{ Time evolution of the density-density correlation function \eqref{rho2}, i.e. the probability density for the measurement of the second particle provided the first particle has been measured at $t=0$ at position $x=0.5$. The measurement breaks the continuous translation symmetry making the probability density nonuniform. During a subsequent evolution, as expressed in the different panels, the density moves along a ring with the period $T=1/2\pi$ but also spreads with the characteristic time increasing with the particle number $N$. All results are obtained for $g_0(N-1)=-15$.
} \label{fig:evo}
\end{figure*}

Eigenstates of the Hamiltonian (\ref{h}) with $\alpha=0$ can be found with the help of the Bethe ansatz \cite{Gaudin}. However, within the Bethe formalism, simulations of the measurement and subsequent time evolution can be in practice performed for a few particles only \cite{syrwid15,syrwid16}. Therefore, we have switched to numerical diagonalization of the Hamiltonian (\ref{h}) with $\alpha=0$ in a truncated Hilbert space. Eigenstates of the system can be written in the Fock states basis, $\prod_{l=l_{\rm min}}^{l_{\rm max}}|n_{l}\rangle$, where $n_l$ denotes number of bosons occupying the single particle mode with momentum $2\pi l$. The values of $l_{\rm min}$ and $l_{\rm max}$ have to be adjusted so that the lowest energy eigenstate $|\psi_0\rangle$ in the subspace corresponding to the CM momentum $P_N$ is properly reproduced. For the value $g_0(N-1)=-15$ we consider here and a central $l=1$ case, seven modes with $l=-2,...,4$ are sufficient to obtain converged results.

\begin{figure}
 \includegraphics[width=1.0\linewidth]{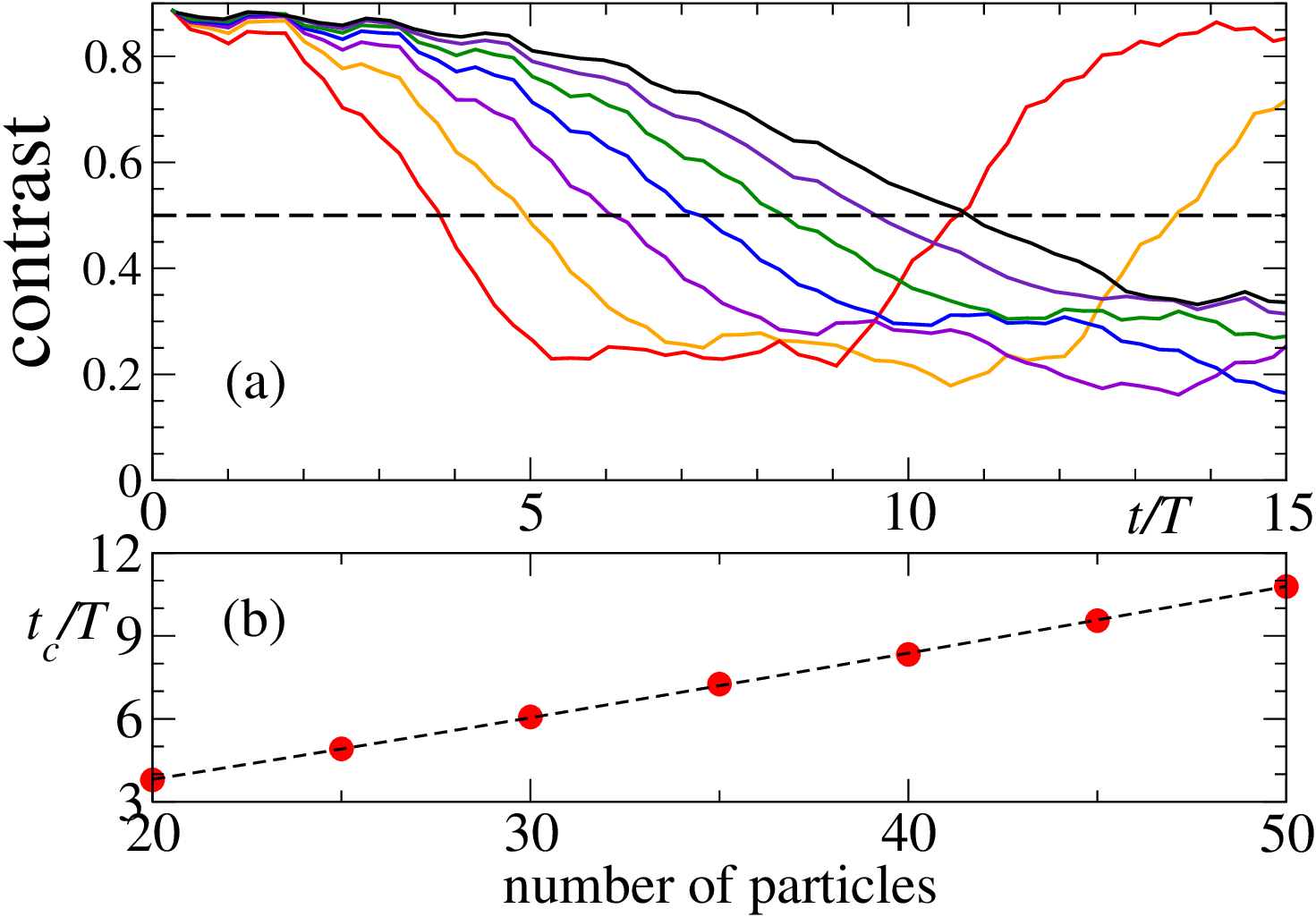}
\caption{(a) The contrast of $\rho_2(x,t)$ as a function of time for different particle numbers, i.e. for $N\in\{20,25,30,35,40,45,50\}$ from left to right. The lifetime $t_c$ plotted in (b) corresponds to the point where the contrast decreases to $C=0.5$. All results are obtained for $g_0(N-1)=-15$.}
 \label{contrast} 
\end{figure}

Probability density for the measurement of a single particle is uniform if the system is prepared in the eigenstate $|\psi_0\rangle$. Assume that at $t=0$, such a measurement results in $x_1=0.5$. Next, we are interested in the probability density $\rho_2(x,t)$ for the position measurement of the second particle at $t>0$ which is related to the density-density correlation function,
\be\label{rho2}
\rho_2(x,t)\propto\langle\psi_0|\hat\psi^\dagger(x,t)\hat\psi(x,t)\;\hat\psi^\dagger(x_1,0)\hat\psi(x_1,0)|\psi_0\rangle,
\ee
where $\hat\psi$'s are the bosonic field operators. This kind of the measurement assumes that at the moment of a detection an atom is removed from the system. Then, in a full analogy to the well-known theory of photon detection, the joint counting rate for two atoms at two moments of time and at two positions is proportional to the second order correlation function of the bosonic field operators \cite{Glauber63,javanainen96}. If $\rho_2(x,t)$ is non-homogeneous in space and reveals periodic evolution that lasts infinitely long in the limit when 
$N\rightarrow\infty$ and $g_0\rightarrow 0$ but $g_0(N-1)=$const., the time crystal behavior is realized. In Fig.~\ref{fig:evo} we show $\rho_2(x,t)$  for different times. Observe a clear breaking of time translational symmetry, i.e. the non-homogeneous probability density for the measurement of the second particle tends to move periodically along the ring with the period $T=1/2\pi$. The time evolution also reveals smearing of the distribution in time. This phenomenon may be observed monitoring the contrast defined as $C(t)=[\max(\rho_2)-\min(\rho_2)]/[\max(\rho_2)+\min(\rho_2)]$ (over the ring, for a fixed $t$) and represented in Fig.~\ref{contrast}(a). The lifetime $t_c$ of the structure created is defined as a point at which $C(t_c)=0.5$. Numerical simulations show that $t_c$ increases linearly with number of particles $N$ as presented in Fig.~\ref{contrast}(b). Thus, in the limit $N\rightarrow\infty$, the symmetry broken state lives forever. For a given number of particles and for sufficiently long time 
evolution the distribution undergoes re-phasing resembling quantum revivals, see Fig.~\ref{contrast}(a).

\begin{figure}
 \includegraphics[width=1.\linewidth]{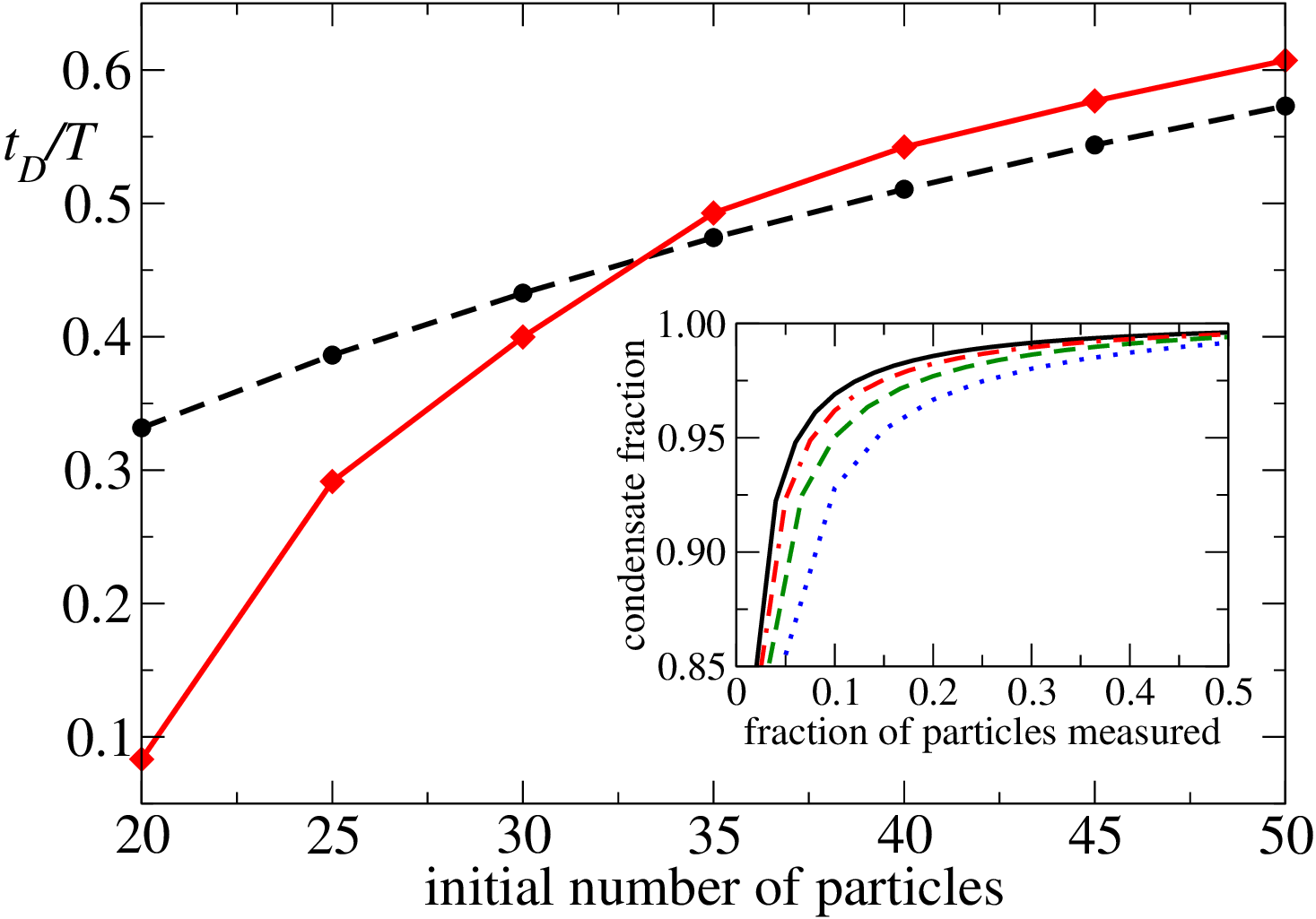}
\caption{The deformation time, $t_D$, of the symmetry broken state (red line with diamonds) as a function of the initial particle number compared with the analytical prediction (black dashed line with circles), see text. At $t=0$, 20\% of particles are measured resulting in a state well approximated by the mean field prediction. The condensate fraction (i.e. the largest eigenvalue of the single particle density matrix) as a function of the fraction of particles measured from the original distribution is presented in the inset for 20, 30, 40, and 50 particles (from right to left). All results are obtained for $g_0(N-1)=-15$.}
\label{fraction}
\end{figure}

It is possible to estimate analytically the time, $t_D$, needed for a noticeable  deformation of the symmetry broken state if we assume that initially we have measured not only one particle but some fraction $\epsilon$ of all particles. Such a measurement drives the system of remaining particles into a Bose-Einstein condensate \cite{javanainen96,dziarmaga03,javanainen08} the quantum state of which can be approximated by a product state $\psi\approx\prod_{i=1}^{N-\epsilon N}\phi(x_i)$ (compare the inset of Fig.~\ref{fraction} where the purification of the condensate due to measurement is presented).  $\phi(x)$ is the largest eigenvalue eigenstate of the reduced single particle density matrix, i.e. the so-called condensate wavefunction.  It is localized around a certain position $q$ on the ring \cite{javanainen08}. The central limit theorem tells us that the probability density for the CM coordinate is $|\chi(x_{\rm CM})|^2\propto e^{-N(1-\epsilon) (x_{\rm CM}-q)^2/\sigma^2}$ where $\sigma^2$ is the mean field variance of the localized 
distribution $|\phi(x)|^2$. Taking 
$\chi(x_{\rm CM})$ as an initial wavefunction of the CM degree of freedom we obtain that its free evolution results in the spreading of the gaussian 
wavepacket, a standard textbook problem. Time when the width of the CM distribution becomes significantly wider than the initial width increases with the number of particles as $N^{1/2}$. 

A direct numerical integration of the exact many-body state confirms our analytical estimates. We first perform the measurement of 20\% of particles (i.e. the measured fraction $\epsilon=0.2$) then time-evolve the resulting many body state and look for the time $t_D$ when the standard deviation of the CM distribution is equal to $\sigma/2$, i.e. half of the width of the Gross-Pitaevskii solution. We have chosen $\sigma/2$ because it is large enough in order to observe deformation of the symmetry broken state and smaller than the length of the ring to avoid the influence of the periodic boundary conditions. For sufficiently large number of particles, $t_D$ closely follows the central limit theorem prediction -- compare the main panel of Fig.~\ref{fraction}. The remaining (small) difference can be attributed to higher order correlations present in the exact many-body state and absent in the  initial mean field state \cite{klaiman16}.
  
Let us consider the possible experimental verification of the results presented. We suggest that the
spontaneous breaking of continuous time translation symmetry can be observed in ultra-cold atomic gases. Tight toroidal atomic trap can mimic a one-dimensional ring. Magnetic flux $\alpha$ can be realized using  methods already known for the  creation of artificial gauge potentials for atoms (see \cite{goldman14} for a recent review), e.g., by inducing rotation of a thermal cloud during the evaporative cooling. In the latter case, the system dissipates to the lowest energy state in the rotating frame and the Coriolis force mimics a magnetic field. The coupling constant $g_0$ can be controlled by means of a Feshbach resonance that allows one to change $s$-wave scattering length of atoms \cite{chin10}. An observation of the spontaneous rotation of a non-uniform atomic density, when the magnetic flux is turned off, will contrast with the same experiment but performed for $g_0(N-1)>-\pi^2$. In the latter case, the corresponding eigenstate of the system is well approximated by a product state of the uniform Gross-
Pitaevskii solution $\phi_0=1$ with no 
position-density signatures. Let us note that we consider neutral bosons in our proposition. Thus the rotating non-uniform atomic density will not radiate and decay as suggested in \cite{bruno13,bruno13b,
else16}.

To summarize, we have analyzed the spontaneous breaking of a continuous time translation symmetry to the discrete symmetry in the time crystal model introduced by Frank Wilczek. If the system is prepared in the ground state, spontaneous rotation of a non-uniform density can not be observed for large number of particles \cite{bruno13,bruno13a}. However, if we start with an excited eigenstate, although the initial single particle density is uniform and does not display any motion, measurement of the position of a single particle reveals a rotation of the remaining particle cloud. The spontaneous rotation that is modeled in the present publication can be observed in ultra-cold atomic gases that allow experimentalists to prepare, control and detect not only many-body ground states but also excited states. A realization of such an experiment will complete an observation of the spontaneous breaking of the continuous time translation symmetry, a major breakthrough as compared with the  spontaneous breaking of a 
discrete time translation symmetry 
already observed in a laboratory in a periodically driven chain of trapped ions \cite{zhang16} or for a driven diamond \cite{choi16}.

%\newpage

%%%%%%%%%%%%%%%%%%%%%%%%%%%%%%%%%%%%%%%%%%%%%%%%%%%%%%%%%%%%%%%%%%%%%%%%%%%%%%%%%%%%%%%
%\begin{figure}
%\includegraphics[width=1.\columnwidth]{fig_exp.eps}
%\caption{}
%\label{dd}
%\end{figure}
%%%%%%%%%%%%%%%%%%%%%%%%%%%%%%%%%%%%%%%%%%%%%%%%%%%%%%%%%%%%%%%%%%%%%%%%%%%%%%%%%%%%%%%

We are grateful to Dominique Delande for fruitful discussions. This work was performed with the support of EU via Horizon2020 FET project QUIC (nr. 641122). We acknowledge support of the National Science Centre, Poland via project No.2016/21/B/ST2/01095 (AS and KS) and 2016/21/B/ST2/01086 (JZ). Moreover, A.S. acknowledges support in the form of a special scholarship from the Marian Smoluchowski Scientific Consortium "Matter-Energy-Future", from the so-called KNOW funding.

\end{document}